\documentclass[notitlepage,a4paper,12pt]{article}
\usepackage{latexsym}
\usepackage{graphicx}
\usepackage{bm}
\begin{document}

\title{\bf\Large{Quantum mechanics of stationary states of particles in external singular spherically and axially symmetric gravitational and electromagnetic fields}}

\date{}

\maketitle

\begin{center}
\author{M. V. Gorbatenko \\ [5pt]
\it Federal State Unitary Enterprise Russian Federal Nuclear Center - All-Russian Research Institute of Experimental Physics, 37 Mira pr., \\ Sarov, Nizhny Novgorod region, 607188, Russia\\
MVGorbatenko@vniief.ru}  \and
\end{center}

\begin{center}
\author{V. P. Neznamov \\ [5pt]
\it Federal State Unitary Enterprise Russian Federal Nuclear Center - All-Russian Research Institute of Experimental Physics, 37 Mira pr., \\ Sarov, Nizhny Novgorod region, 607188, Russia\\
VPNeznamov@vniief.ru, vpneznamov@mail.ru}
\end{center}

%\maketitle

\begin{abstract}
The report considers the interaction of scalar particles, photons and
fermions with the gravitational and electromagnetic Schwarzschild, Reissner-Nordstr\"{o}m,
Kerr and Kerr-Newman fields. The behavior of effective potentials in the relativistic
Schr\"{o}dinger-type second-order equations is analyzed. It was found that
the quantum theory is incompatible with the hypothesis of the existence of
classical black holes with event horizons of zero thickness that were
predicted based on solutions of the general relativity (GR) with zero and
non-zero cosmological constant $\Lambda$. The alternative may be presented
by compound systems, i.e., collapsars with fermions in stationary bound
states.\\ [8pt]
[8pt]
 
\end{abstract}

\section{Introduction}	

In the quantum mechanics, G.T.Horowitz and D.Marolf in Ref. \cite{1} actually proposed a quantum mechanical hypothesis of cosmic censorship. They
write in the Introduction: ''We will say that a system is nonsingular, when
the evolution of any state is uniquely defined for all time. If this is not
the case, then there is some loss of predictability and we will say that the
system is singular''. Similarly to Penrose in Ref. \cite{2}, we should add that such singular systems cannot exist  in nature.

Let us present some numerical characteristics of singular and nonsingular
systems. For the radial second-order equations reduced to the
Schr\"{o}dinger-type equations with the effective potentials $U_{eff} \left(
r \right)$, the behavior of effective potentials in the neighborhood of the black hole event
horizons has the form of an infinitely deep potential well
\[
\left. {U_{eff} \left( r \right)} \right|_{r \to r_{\pm } }
=-\,{K_{1} } /{\left( {r -r_{\pm}} \right)^{2}}.
\]
Here $r_{\pm}$ are radii of external and inner event horizons.

If $K_{1} \geq 1/8$, a so-called ''fall'' of a particle to the
event horizon occurs \cite{3,4}. In this case the system is singular. The behavior of
the radial function of the Schr\"{o}dinger-type equation has the following
form
\[
\left. {R\left( r \right)} \right|_{r \to r_{\pm } } \sim \left(
{r -r_{\pm } } \right)^{1/2}\sin \left( {\sqrt {K_{2} } \ln \left( {r
-r_{\pm } } \right)+\delta } \right),
\]
where $K_{2}=2\left( K_{1}-1/8\right) $.

As  $r \to r_{\pm }$, the functions of stationary states of discrete and continuous spectrum $R\left( r \right)$ have an unlimited number of zeros, discrete levels of energy ''dive'' into the area of negative continuum. The function  $R\left( r \right)$ does not have defined values at $r=r_{\pm}$.

Also, the system will be singular, if the exponent of the denominator in the
expression for the effective potential higher than two. In this case
\[
\left. {R_{s} \left( r \right)} \right|_{r \to r_{\pm } } \sim
\left( {r - r_{\pm } } \right)^{s/4}\sin \left( {\frac{2}{s-2}\sqrt
{\frac{2K_{1} }{\left( {r -r_{\pm } } \right)^{s-2}}} +\delta_{s} }
\right),
\]
where $\delta ,\,\,\delta_{s} $ are arbitrary phases, $s>2$ is the exponent in
the effective potential.

In the Hamiltonian formulation, the mode of a ``particle fall'' to the event horizons means
that the Hamiltonian $H$ has nonzero deficiency indexes \cite{5}-\cite{7}. For elimination of this mode, it is necessary to choose the additional boundary conditions on the event horizons. The self-conjugate extension of the Hermitian operator $H$ is determined by this choice.

In the quantum theory there is an example confirming the quantum mechanical hypothesis of cosmic censorship. This is ``$Z>137$ catastrophe''.

Let us consider solutions of the Schr\"{o}dinger-type equation with the effective potential for fermions in the Coulomb field $V\left( \rho \right)=-Z\alpha_{fs}/\rho $. The asymptotes of the effective potential as $\rho\rightarrow 0$ has the form

\[
\left. U^{C}_{eff} \right|_{\rho\rightarrow 0}=-\frac{\left( Z \alpha_{fs} \right)^{2} - 3/4 + \left( 1-\kappa^{2}\right) }{2\rho^{2}}.
\]

In the asymptotes, dependently on $Z$ it is possible to single out three characteristic areas. For example, we consider these areas for the bound states $1S_{1/2} \left( \kappa=-1 \right), 2P_{1/2} \left( \kappa=+1 \right) $. In the first ares $1\leq Z <\sqrt{3}/2\alpha_{fs}$ as $\rho\rightarrow 0$ there exists the positive barrier $\sim 1/\rho^{2}$ with the following potential well. As $Z=Z_{cr}=\sqrt{3}/2\alpha_{fs}\approx 118.7$  the potential barrier disappears; as for $Z>Z_{cr}$ as $\rho\rightarrow 0$ the potential well $-K/\rho^{2}$ remains. In the second area $119\leq Z<137$ the coefficient is $K<1/8$, which permits the existence of the fermion stationary bound states. In the third area $Z\geq137$ as $\rho\rightarrow 0$ there is the potential well with $K\geq1/8$, which is indicative of the implementation of the mode of ``a fall'' to the center. For elimination of this mode, it was offered to take into account the finite sizes of nuclei \cite{8,9}. The system of ''an electron in the Coulomb field of the finite-size atomic nucleus'' is nonsingular.

\section{Effective Potentials}

For a closed system of ''a particle in the external force field'', quantum
mechanics allows existence of stationary states with real energies. The stationary states involve both the states of discrete spectrum (bound states) and states of continuous spectrum (scattering states).
In this case, the wave function of the particle is written as
\[
\psi \left( {{\rm {\bf r}},t} \right)=\psi \left( {{\rm {\bf r}}}
\right)e^{-iEt},
\]
where $E$ is a real particle energy. Here and hereafter, we use the system of units $\hbar =c=1$.

We will explore in closed systems the existence possibility of stationary
states of scalar particles $\left( {S=0} \right)$, photon
$\left( {S=1} \right)$, fermions$\left( {S=1/2} \right)$ interacting with the Schwarzschild,
Reissner-Nordstr\"{o}m, Kerr and Kerr-Newman black holes.

Scheme of our analysis is the following:

\begin{enumerate}
\item Scalar massive particles with charge $q \left( {S=0} \right)$.
\begin{enumerate}
\item Klein-Gordon equation
\[
\left( {-g} \right)^{-1/2} \left( \frac{\partial }{\partial x^{\mu }}-iq A_{\mu} \right)  \left[ {\left( {-g} \right)^{1/2}g^{\mu \nu } \left( \frac{\partial }{\partial x^{\nu }}-iq A_{\nu} \right) \Phi } \right]+m^{2}\Phi =0, \]

where $g$ are the metrics determinant, $A_{\mu}$ are potentials of electromagnetic field.

\item Separation of variables
\[
\Phi \left( {{\rm {\bf r}},t} \right)=\sum\limits_{l,m_{\varphi } } R_{l}^{S,RN}
\left( r \right)Y_{lm_{\varphi } } \left( {\theta ,\varphi }
\right)e^{-iEt} \]
(for the spherically symmetric Schwarzschild,
Reissner-Nordstr\"{o}m metrics, $Y_{lm_{\varphi}}  \left( \theta ,\varphi \right)$ is spherical harmonics),
\[
\Phi_{KN} \left( {{\rm {\bf r}},t} \right)=\sum\limits_{l,\,m_{\varphi } }
{R_{l}^{K,KN} \left( r \right)S\left( \theta \right)e^{-iEt}} e^{i\,m_{\varphi }
\,\varphi }\]
 (for the axially symmetric Kerr, Kerr-Newman metrics, $S\left(
\theta \right)$ are the oblate spheroidal harmonic functions $S_{lm_{\varphi
} } \left( {ic\,\,\cos \theta } \right)$, where $c=a^{2}\left( {\left(
{{E^{2}}/{m^{2}}} \right)-1} \right)$) \cite{10}.
\item Second-order equation for radial functions $R_{l} \left( r \right)$
\[
\frac{d^{2}R_{l} }{d r^{2}}+A\left( r \right)\frac{dR_{l} }{d r
}+B\left( r \right)R_{l} =0,\]

Here, for brevity, the functions $R_{l} \left(  r \right) $ for different metrics are denoted identically.
\item Reduction to form of the relativistic Schr\"{o}dinger equation with the effective potential $U_{eff} \left( r \right)$
\[
\bar{{R}}_{l} \left( r \right)=R_{l} \left( r \right)\exp
\frac{1}{2}\int {A\left( {{r }'} \right)} d{r }',\]

\[
\frac{d^{2}\bar{{R}}_{l} \left( r \right)}{d r^{2}}+2\left( {E_{Schr}
-U_{eff} \left( r \right)} \right)\bar{{R}}_{l} \left( r \right)=0,\]

\[
U_{eff} \left( r \right)=E_{Schr} +\frac{1}{4}\frac{dA}{dr
}+\frac{1}{8}A^{2}-\frac{1}{2}B,\]

\[
E_{Schr} =\left( E^{2}-m^{2} \right)/2.\]

\item Study of behavior of effective potential in the neighborhood of the event horizons.	
\end{enumerate}
\item Photon $\left( S=1 \right) $.

The Maxwell equations for the spherically symmetric Schwarzschild and  Reissner-Nordstrom metrics were written in three-dimensional form (in three-dimensional space with metric $\gamma_{ik}=- g_{ik} + \left(  g_{0i} g_{0k}/g_{00} \right)  $) \cite{11}. Separation of the variables was performed via expansion of electric ${\bf E} \left( {\bf x}, t \right) $  and magnetic ${\bf H} \left( {\bf x}, t \right) $  fields in terms of the vector harmonics of the electric, magnetic and longitudinal types. For the axially symmetric Kerr, Kerr-Newman metrics, the variables separation procedure of O.Lunin was used \cite {12}.
As a result, the real effective potentials of the Schrodinger-type relativistic equations were obtained.
Earlier, for the Kerr metric, S.Chandrasekhar in Ref. \cite{Chandrasekhar} used Tukolsky's method (see Ref. \cite{Teukolsky}-\cite{Teukolsky1}) at separation of the variables in the Maxwell equations. This method describes only the states with ''electrical'' polarization (see Ref. \cite{12}) and leads to the second-order equations for the radial wave functions with the complex wave potentials and complex energies. According to S.Chandrasekhar in Ref. \cite{Chandrasekhar}, the singular solutions of these equations on the external event horizon can be made regular by changing the basis.

\item Fermion with charge $q \left( S=1/2 \right) $.	

For metrics under consideration, the effective potentials of the Schr\"{o}dinger-type equation were obtained in Refs. \cite{13} - \cite{15}. These references also contain the solutions for the stationary bound states with energies $E^{st}$.
\end{enumerate}

\section{The Behavior of Effective Potentials in\\ the Neighborhood of the Event
Horizons}

As the result, we obtained the following leading singularities of the effective potentials $U_{eff}(r)$ in the neighborhoods of event horizons:

\begin{enumerate}
\item Schwarzschild field.
\begin{enumerate}
\item Scalar particle, photon, fermion with $E\ne E^{st} $
\[
\left. {U_{eff} } \right|_{r\to r_{0} } =-\frac{1}{\left( {r-r_{0} }
\right)^{2}}\left( {\frac{1}{8}+\frac{r_{0}^{2} E^{2}}{2}} \right),
\]
\item Fermion with $E=E^{st} =0$
\[
\left. {U_{eff} } \right|_{r\to r_{0} } =-\frac{3}{32}\frac{1}{\left(
{r-r_{0} } \right)^{2}},
\]	
\end{enumerate}
\item Reissner-Nordstr\"{o}m field.
\begin{enumerate}
\item Scalar particle, fermion with $E\ne E^{st} $
\[
\left. {U_{eff} } \right|_{r \to \left( {r_{\pm } } \right)_{RN} }
=-\,\,\frac{1}{\left( {r -\left( {r_{\pm } } \right)_{RN} }
\right)^{2}}\left[ {\frac{1}{8}+\frac{\left( {E -\frac{qQ}{\left( {r_{\pm } } \right)_{RN} }} \right)^{2}\left( {r_{\pm
} } \right)_{RN}^{4} }{2\left[ {\left( {\left( {r_{+} } \right)_{RN}
-\left( {r_{-} } \right)_{RN} } \right)} \right]^{2}}} \right].
\]
\item Photon
\[
\left. {U_{eff} } \right|_{r\to
\left( {r_{\pm } } \right)_{RN} } =-\,\,\frac{1}{\left( {r-\left( {r_{\pm }
} \right)_{RN} } \right)^{2}}\left[ {\frac{1}{8}+\frac{E^{2}\left(
{r_{\pm } } \right)_{RN}^{4} }{2\left[ {\left( {r_{+} } \right)_{RN} -\left(
{r_{-} } \right)_{RN} } \right]^{2}}} \right].
\]
\item Fermion with $E=E^{st} ={qQ} \mathord{\left/ {\vphantom {{qQ}
{\left( {r_{\pm } } \right)_{RN} }}} \right. \kern-\nulldelimiterspace}
{\left( {r_{\pm } } \right)_{RN} }$

\[
\left. {U_{eff} }
\right|_{r \to \left( {r_{\pm } } \right)_{RN} }
=-\frac{3}{32}\frac{1}{\left( {r -\left( {r_{\pm } } \right)_{RN} }
\right)^{2}}.
\]
\end{enumerate}
\item Kerr, Kerr-Newman fields.
\begin{enumerate}
\item Scalar particle, fermion with $E \ne E^{st} $
\[
\left. {U_{eff}^{KN} } \right|_{r \to \left( {r_{\pm } } \right)_{KN}
} =-\frac{1}{\left( {r -\left( {r_{\pm } } \right)_{KN} }
\right)^{2}}\left[ {\frac{1}{8}+\frac{\left( {E -E
^{st} } \right)^{2}\left( {\left( {r_{\pm } } \right)_{KN}^{2}
+a^{2} } \right)^{2}}{2\left[ {\left( {r_{+} } \right)_{KN}
-\left( {r_{-} } \right)_{KN} } \right]^{2}}} \right].
\]
\item Photon
\[
\left. {U_{eff} } \right|_{r\to
\left( {r_{\pm } } \right)_{KN} } =-\,\,\frac{1}{\left( {r-\left( {r_{\pm }
} \right)_{KN} } \right)^{2}}\left[ {\frac{1}{8}+\frac{\left( {E -\frac{am_{\varphi}}{\left( {r_{\pm } } \right)_{KN}^{2} + a^{2} }} \right)^{2} \left( {\left( {r_{\pm } } \right)_{KN}^{2}
+a^{2} } \right)^{2}}{2\left[ {\left( {r_{+} } \right)_{KN} -\left(
{r_{-} } \right)_{KN} } \right]^{2}}} \right].
\]
\item Fermion with $E=E^{st} =\frac{\alpha m_{\varphi} +Qq \left( r_{\pm} \right)_{KN} }{\alpha^{2}+ \left( r_{\pm} \right)^{2}_{KN} }$
\[
\left. {U_{eff}^{KN} } \right|_{r \to \left( {r_{\pm } } \right)_{KN}
} =-\frac{3}{32}\frac{1}{\left( {r -\left( {r_{\pm } } \right)_{KN} }
\right)^{2}}.
\]	
\end{enumerate}
\end{enumerate}

Here, $r_{0}=2GM$ is the gravitational radius of the Schwarzschild field, $\left( r_{\pm}\right)_{RN} $, $\left( r_{\pm}\right)_{KN} $ are the external and inner radii of the event horizons of the Reissner-Nordstr\"{o}m, Kerr, Kerr-Newman fields, $Q$ is the charge of the Reissner-Nordstr\"{o}m field source, $a=J/M$, where $J$ is the angular momentum of the source of the Kerr and Kerr-Newman fields.

\subsection{Non-zero cosmological constant $\Lambda $}

We have proved that Kerr-Newman-(A)dS, Kerr-(A)dS,
Reissner-Nordstr\"{o}m-(A)dS and Schwarzschild-(A)dS black holes have the
same divergences of effective potentials near the event
horizons as in case of $\Lambda =0$. For example, for the most general
Kerr-Newman-dS field $\left( {\Lambda >0}
\right)$ as $r\to r_{+} $:

\begin{itemize}
\item Scalar particle, fermion with $\Omega_{+} \ne 0$
\[
\left. {U_{eff}^{KN-dS} } \right|_{r\to r_{+} } =-\,\frac{1}{\left( {r-r_{+}
} \right)^{2}}\left\{ {\frac{1}{8}+\frac{\Omega_{+}^{2} }{2\left[ {\left(
{r_{+} -r_{-} } \right)\left( {r_{+} -r_{\Lambda }^{+} } \right)\left(
{r_{+} -r_{\Lambda }^{-} } \right)} \right]^{2}}} \right\},
\] \\
where  $\Omega_{+}
=\Xi \left( {E\left( {r^{2}+a^{2}} \right)-m_{\varphi } a-\frac{qQr}{\Xi }}
\right),
\Xi =1+a^{2}\frac{\Lambda }{3}$, $r_{\Lambda }^{+}$ is the cosmological horizon.\\

\item Photon
\[
\left. {U_{eff}^{KN-dS} } \right|_{r\to r_{+} }-\,\frac{1}{\left( {r-r_{+} } \right)^{2}}\left\{
{\frac{1}{8}+\frac{\left( {\omega \left( {r^{2}+a^{2}} \right)-m_{\varphi }
a} \right)^{2}}{2\left[ {\left( {r_{+} -r_{-} } \right)\left( {r_{+}
-r_{\Lambda }^{+} } \right)\left( {r_{+} -r_{\Lambda }^{-} } \right)}
\right]^{2}}} \right\}.
\]
\end{itemize}

We also proved that Kerr-Newman-AdS black hole in the minimal five-dimensional gauged
supergravity has the same divergence of effective potentials
near the event horizons as in the four-dimensional case.

\subsection{Coordinate transformations of the Schwarzschild metric}

For the scalar particles, it is proved that transition to the
Eddington-Finkelstein and the Painlev\'{e}-Gullstrand metrics preserves the
leading singularity in the neighborhood of event horizon at the real
axis.

For fermions, for the Schwarzschild space-time in the isotropic coordinates,
as well as the Eddington-Finkelstein, Painlev\'{e}-Gullstrand,
Lemaitre-Finkelstein and Kruskal-Szekeres metrics, it is proved that in case
of the stationary bound state $E^{st} =0$ the leading
singularity in the event horizon neighborhood is also preserved \cite{16}.

\section{Discussion}

The analysis shows that in the considered
gravitational fields the existence of the stationary particle states is
impossible. Systems of ``a particle in the gravitational and electromagnetic
fields'' are singular.

For all the considered metrics and particles with different spins it is
characteristic the universal nature of the divergence of the effective
potentials near to the event horizons. Uncovered singularities doesn't allow
using the quantum theory in full measure that leads to the necessity of
changing the initial formulation of the physical problem.

\section{Is it Possible to Use the Concerned GR Solutions from the Standpoint of
Quantum Mechanics? Is it Possible to Blur the Event Horizons in a Natural
Way?}

Our answer to these questions is positive and we propose to supplement the
gravitational collapse mechanism.

Let us, at the last collapse stage the gravitational field captures the
half-spin particles, which after the formation of the event horizons will
get into the stationary bound states with $E=E^{st}$ both under the inner and above
external event horizons. The bound particles with $E=E^{st}$ are, with overwhelming probability, at the distance from the event horizon within the range from zero to several fractions of the Compton wave length of a fermion (see Refs. \cite{13} - \cite{15} and Fig. 1). For the next particles interacting with such
compound systems, the self-consistent gravitational and electromagnetic
fields will be determined both by the collapsar mass and charge, and by the
masses and charges of the fermions in the stationary bound states with
$E=E^{st}$ in the neighborhood of the event horizons.

Obviously, such system may be nonsingular. To provide a strong proof, there
are required exact calculations of the self-consistent gravitational and
electromagnetic fields, as well as the proof that the stationary states of
the quantum-mechanical probe particles exist in these fields.

\begin{figure}[h]
\centerline{\includegraphics[width=10cm]{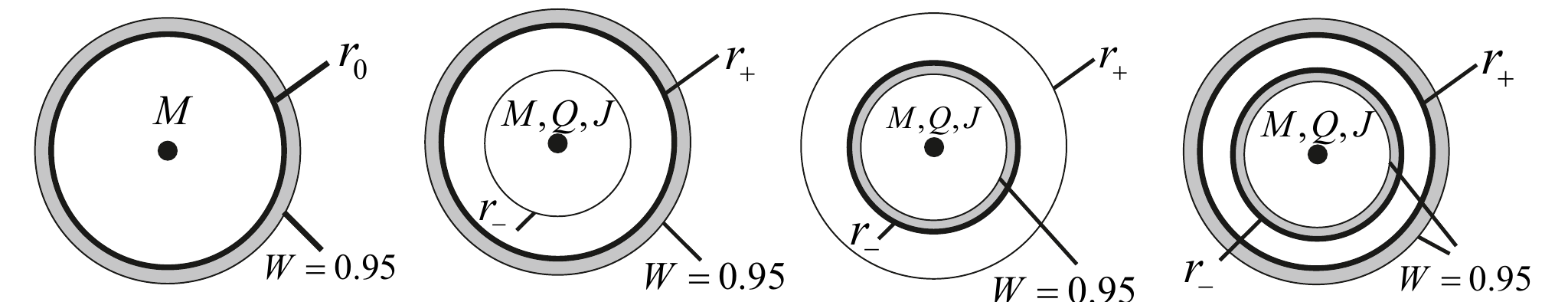}}
\caption{The compound systems: collapsar $+$ fermion. $W$ is the integral  probability of fermion detection. \label{f1}}
\end{figure}

\section{Conclusion}
The quantum theory is incompatible with the existence of the Schwarzschild,
Reissner-Nordstr\"{o}m, Kerr, Kerr-Newman black holes with event horizons of
zero thickness that were predicted based on the GR solutions with zero and
non-zero cosmological constant $\Lambda $.


\begin{thebibliography}{99}    %for 1 digit

\bibitem{1} G. T. Horowitz and D. Marolf, {\it Phys. Rev. D} {\bf 52}, 5670 (1995).

\bibitem{2} R. Penrose, {\it Rivista del Nuovo Cimento}, Serie I, 1, Numero Speciale: {\bf 252} (1969).

\bibitem{3} L. D. Landau and E. M. Lifshitz, {\it Quantum Mechanics. Nonrelativistic Theory} (Pergamon Press, Oxford, 1965).

\bibitem{4} A. M. Perelomov and V. S. Popov, {\it Theor. Math. Phys.}  {\bf 4}, 664 (1970).

\bibitem{5} K. Meetz, {\it Nuovo Cim.} {\bf 34}, 690 (1964).

\bibitem{6} H. Behnce, {\it Nuovo Cim. A} {\bf 55}, 780 (1968).

\bibitem{7} A. Wightman, Introduction to some aspects of the relativistic dynamics of quantized fields, in{\it 1964 Carg\`{e}se Summer School Lectures}, ed. M.L\'{e}vy (Gordon and Breach, New York, 1967), p.171.

\bibitem{8} I. Ya. Pomeranchuk and Y. A. Smorodinsky, {\it Jour. Phys. USSR} {\bf 9}, 97 (1945).

\bibitem {9} W. Paper and W. Griener, {\it Zs. Phys.} {\bf 218}, 327 (1969).

\bibitem{10} V. B. Bezerra, H. S. Vieira and Andr\'{e} A. Costa, {\it Class. Quantum Grav.} {\bf 31}, 045003 (2014).

\bibitem{11} L. D. Landau and E. M. Lifshitz, {\it The Classical Theory of Fields} (Pergamon Press, Oxford, 1975).

\bibitem{12} O. Lunin, {\it J. High Energ. Phys.}  {\bf 2017}, 138 (2017).

\bibitem{Chandrasekhar} S. Chandrasekhar,{\it The Mathematical Theory of Black Holes} (Oxford University Press, 1983).

\bibitem {Teukolsky} S. A. Teukolsky, {\it Phys. Rev. Lett.} {\bf 29}, 1114 (1972); S. A. Teukolsky, {\it Astrophys. J.} {\bf 185}, 635 (1973).

\bibitem {Teukolsky1} S. A. Teukolsky, {\it Astrophys. J.} {\bf 185}, 635 (1973).

\bibitem{13} V. P. Neznamov and I. I. Safronov, {\it J. Exp. Theor. Phys.} {\bf 127}, 647 (2018).

\bibitem{14} V. P. Neznamov, I.I.Safronov and V. E. Shemarulin, {\it J. Exp. Theor. Phys.} {\bf 127}, 684 (2018).

\bibitem{15} V. P. Neznamov, I.I.Safronov and V. E. Shemarulin, {\it J. Exp. Theor. Phys.} {\bf 128}, 64 (2019).

\bibitem {16} M. V. Gorbatenko and V. P. Neznamov, {\it Theor. Math. Phys.}  {\bf  198}, 425 (2019).


\end{thebibliography}
\end{document}